\documentclass[]{article}
\usepackage{latexsym}
\usepackage{amssymb}
\usepackage{amsmath}
\usepackage{amscd}
\usepackage{amsthm}
\usepackage[left=2cm,top=2.5cm,right=2.5cm,bottom=1.5cm]{geometry}

\usepackage[dvips]{graphicx}
\usepackage{hyperref}
\begin{document}
\begin{center}
\Large{\bf{String Cosmology in Anisotropic Bianchi-II Space-time}}
\\
\vspace{10mm} \normalsize{Suresh Kumar}\\ \vspace{5mm} \normalsize{
Department of Applied Mathematics,\\ Delhi Technological University
(Formerly Delhi College of Engineering),  \\Bawana Road, Delhi-110
042, India.} \normalsize{E-mail: sukuyd@gmail.com}
\end{center}

\begin{abstract}
The present study deals with a spatially homogeneous and anisotropic
Bianchi-II cosmological model representing massive strings.  The
energy-momentum tensor, as formulated by Letelier (1983), has been
used to construct a massive string cosmological model for which the
expansion scalar is proportional to one of the components of shear
tensor. The Einstein's field equations have been solved by applying
a variation law for generalized Hubble's parameter that yields a
constant value of deceleration parameter in Bianchi-II space-time. A
comparative study of accelerating and decelerating modes of the
evolution of universe has been carried out in the presence of string
scenario. The study reveals that massive strings dominate the early
Universe. The strings eventually disappear from the Universe for
sufficiently large times, which is in agreement with the current
astronomical observations.
\end{abstract}
\smallskip

Keywords: Massive string, Bianchi-II model, Accelerating universe

PACS number: 98.80.Cq, 04.20.-q, 04.20.Jb
\section{Introduction}
In recent years, there has been considerable interest in string
cosmology. Cosmic strings are topologically stable objects which
might be found during a phase transition in the early universe
(Kibble \cite{ref1}). Cosmic strings play an important role in the
study of the early Universe. These arise during the phase transition
after the big bang explosion as the temperature goes down below some
critical temperature as predicted by grand unified theories
(Zel'dovich et al. \cite{ref2}; Kibble \cite{ref1,ref3}; Everett
\cite{ref4}; Vilenkin \cite{ref5}). It is believed that cosmic
strings give rise to density perturbations which lead to the
formation of galaxies (Zel'dovich \cite {ref6}). However, recent
observations suggest that cosmic strings cannot be wholly
responsible for either the CMB fluctuations or the observed
clustering of galaxies \cite {ref6a,ref6b}. The cosmic strings have
stress-energy, and couple to the gravitational field. Therefore, it
is interesting to study the gravitational effects that arise from
strings. The pioneering work in the formulation of the
energy-momentum tensor for classical massive strings was done by
Letelier \cite{ref7} who considered the massive strings to be formed
by geometric strings with particle attached along its extension.
Letelier \cite{ref8} first used this idea in obtaining cosmological
solutions in Bianchi-I and Kantowski-Sachs space-times. Stachel
\cite{ref9} has studied massive string.

The present day Universe is satisfactorily described by homogeneous
and isotropic models given by the FRW space-time. But at smaller
scales, the Universe is neither homogeneous and isotropic nor do we
expect the Universe in its early stages to have these properties.
Homogeneous and anisotropic cosmological models have been widely
studied in the framework of general relativity in the search of a
realistic picture of the Universe in its early stages. Although
these are more restricted than the inhomogeneous models which
explain a number of observed phenomena quite satisfactorily. A
spatially homogeneous Bianchi model necessarily has a
three-dimensional group, which acts simply transitively on
space-like three-dimensional orbits. Here we confine ourselves to
models of Bianchi-II. Asseo and Sol \cite{ref41} emphasized the
importance of Bianchi type-II Universe. Bianchi type-II space-time
has a fundamental role in constructing cosmological models suitable
for describing the early stages of evolution of Universe.

Roy and Banerjee \cite{ref42} have dealt with locally rotationally
symmetric (LRS) cosmological models of Bianchi type-II representing
clouds of geometrical as well as massive strings. Wang \cite{ref43}
studied the Letelier model in the context of LRS Bianchi type-II
space-time. Recently, Pradhan et al. \cite{ref46,ref47} and
Amirhashchi and Zainuddin \cite{ref48} obtained LRS Bianchi type II
cosmological models with perfect fluid distribution of matter and
string dust, respectively. Belinchon \cite{ref49,ref50} studied
Bianchi type-II space-time in connection with massive cosmic string
and perfect fluid models with time varying constants under the
self-similarity approach respectively. Recently, Tyagi and Sharma
\cite{ref54} have investigated string cosmological models in Bianchi
type-II space-time.

Motivated by the above discussions, in this paper, we have
investigated a new class of Bianchi type-II cosmological models for
a cloud of strings by using the law of variation for generalized
mean Hubble's parameter. This approach is different from what the
other authors have adapted. The paper is organized as follows. The
metric and the field equations are presented in Section 2. Section 3
deals with exact solutions of the field equations with cloud of
strings. Physical behavior of the derived model is elaborated in
detail. Finally, in Section 4, concluding remarks are given.

\section{The metric and field  equations}
We consider totally anisotropic Bianchi type-II line element, given by
\begin{equation}
\label{eq1} ds^{2} = - dt^{2} + A^{2}(dx - zdy)^{2} + B^{2} dy^{2} +
C^{2} dz^{2},
\end{equation}
where the metric potentials $A$, $B$ and $C$ are functions of $t$
alone. This ensures that the model is spatially homogeneous.

The Einstein's field equations ( in gravitational units $c = 1, 8\pi
G = 1 $) read as
\begin{equation}\label{eq1a}
 G^{i}_{j}=  -
T^{i}_{j},
\end{equation}
where $G^{i}_{j}= R^{i}_{j} - \frac{1}{2} R g^{i}_{j}$ is the
Einstein tensor. The energy-momentum tensor $T^{i}_{j}$ for a cloud
of massive strings and perfect fluid distribution is taken as
\begin{equation}
\label{eq2} T^{i}_{j} = (\rho + p)v^{i}v_{j} + p g^{i}_{j} -\lambda
x^{i}x_{j},
\end{equation}
where $p$ is the isotropic pressure; $\rho$ is the proper energy density for a cloud strings with particles
attached to them; $\lambda$ is the string tension density; $v^{i}=(0,0,0,1)$ is the four-velocity of the
particles, and $x^{i}$ is a unit space-like vector representing the direction of string. The vectors $v^{i}$
and $x^{i}$ satisfy the conditions
\begin{equation}
\label{eq3} v_{i}v^{i}=-x_{i}x^{i}=-1,\;\; v^{i}x_{i}=0.
\end{equation}

Choosing $x^{i}$ parallel to $\partial/\partial z$, we have
\begin{equation}
\label{eq4} x^{i} = (0,0,C^{-1},0).
\end{equation}

Here the cosmic string has been directed along z-direction in order
to satisfy the condition $T^{1}_{1}=T^{2}_{2}$. As a result, the
off-diagonal component of Einstein tensor, viz.,
$G^{1}_{2}=z(T^{2}_{2}-T^{1}_{1})$ vanishes. A detailed analysis
about the choice of energy-momentum tensor for Bianchi type-II
models is given by Saha \cite{saha}.

If the particle density of the configuration is denoted by
$\rho_{p}$, then
\begin{equation}
\label{eq5} \rho = \rho_{p}+\lambda.
\end{equation}

The Einstein's field equations (\ref{eq1a}) for line element
(\ref{eq1}) with energy-momentum tensor (\ref{eq2}), lead to the
following set of independent differential equations:
\begin{equation}
\label{eq7} \frac{\ddot{B}}{B} + \frac{\ddot{C}}{C} +
\frac{\dot{B}\dot{C}}{BC} - \frac{3}{4}\frac{A^{2}} {B^{2}C^{2}} =
-p \;,
\end{equation}
\begin{equation}
\label{eq8} \frac{\ddot{C}}{C} + \frac{\ddot{A}}{A} + \frac{\dot{C}\dot{A}}{CA} + \frac{1}{4}\frac{A^{2}}
{B^{2}C^{2}} = -p\;,
\end{equation}
\begin{equation}
\label{eq9} \frac{\ddot{A}}{A} + \frac{\ddot{B}}{B} +
\frac{\dot{A}\dot{B}}{AB} + \frac{1}{4}\frac{A^{2}} {B^{2}C^{2}} =
-p+ \lambda\;,
\end{equation}
\begin{equation}
\label{eq10} \frac{\dot{A}\dot{B}}{AB} + \frac{\dot{B}\dot{C}}{BC} + \frac{\dot{C}\dot{A}}{CA} -
\frac{1}{4}\frac{A^{2}}{B^{2}C^{2}} = \rho\;.
\end{equation}

Here, and in what follows, an over dot indicates ordinary
differentiation with respect to $t$. The energy conservation
equation $\;T^{ij}_{\;\;\;;j}=0$, leads to the following expression:
\begin{equation}
\label{eq11} \dot\rho + (\rho + p)\left(\frac{\dot{A}}{A} +
\frac{\dot{B}}{B} + \frac{\dot{C}}{C}\right) -
\lambda\frac{\dot{C}}{C} = 0\;,
\end{equation}
which is a consequence of the field equations (\ref{eq7})-(\ref{eq10}).
\section{Solutions of the Field Equations}
Equations (\ref{eq7})-(\ref{eq10}) are four equations in six unknown
parameters $A$, $B$, $C$, $p$, $\rho$ and $\lambda$. Two additional
constraints relating these parameters are required to obtain
explicit solutions of the system.

First, we utilize the special law of variation for the Hubble's
parameter given by Berman \cite{ref59}, which yields a constant
value of deceleration parameter. Here, the law reads as
\begin{equation}
\label{eq13} H = \ell (ABC)^{-\frac{n}{3}},
\end{equation}
where  $\ell>0$  and  $n\geq0$ are constants. Such type of relations
have already been considered by Berman and Gomide \cite{ref60} for
solving FRW models. Later on, many authors (see, Kumar and Singh
\cite{ref63}, Akarsu and Kilinc \cite{ref64} and references therein)
have studied flat FRW and Bianchi type models by using the special
law for Hubble's parameter that yields constant value of
deceleration parameter.

Considering $(ABC)^{\frac{1}{3}}$ as the average scale factor of the
anisotropic Bianchi-II space-time, the average Hubble's parameter
may be written as
\begin{equation}
\label{eq14} H = \frac{1}{3}\left(\frac{\dot{A}}{A} +
\frac{\dot{B}}{B} + \frac{\dot{C}}{C}\right).
\end{equation}

Equating the right hand sides of (\ref{eq13}) and (\ref{eq14}), and
integrating, we obtain
\begin{equation}
\label{eq16} ABC = (n \ell t + c_{1})^{\frac{3}{n}}, \;\;\;\;(n\neq
0)
\end{equation}
\begin{equation}
\label{eq17} ABC = c_{2}^{3}e^{3 \ell t},\;\;\;\;(n=0)
\end{equation}
where $c_{1}$ and $c_{2}$ are constants of integration. Thus, the
law (\ref{eq13}) provides power-law (\ref{eq16}) and exponential-law
(\ref{eq17}) of expansion of the Universe.

Following Pradhan and Chouhan \cite{ref55a}, we assume that the
component $\sigma^{1}_{~1}$ of the shear tensor $\sigma^{j}_{~i}$ is
proportional to the expansion scalar ($\theta$). This condition
leads to the following relation between the metric potentials:
\begin{equation}
\label{eq12} A = (BC)^{m},
\end{equation}
where $m$ is a positive constant.

Now, subtracting (\ref{eq8}) from (\ref{eq7}), and taking integral
of the resulting equation two times, we get
\begin{equation}
\label{eq19} \frac{B}{A} = c_{4} \exp \left[\int
\left\{\frac{1}{ABC}\int\frac{ A^{3}}{BC}dt\right\} dt+c_{3} \int
\frac{1}{ABC} dt \right],
\end{equation}
where  $\;c_{3} $ and  $\;c_{4} $ are constants of integration. In
the following subsections, we discuss the string cosmology using the
power-law (\ref{eq16}) and exponential-law (\ref{eq17}) of expansion
of the Universe.
\subsection{String Cosmology with Power-law}
Solving the equations (\ref{eq16}), (\ref{eq12}) and (\ref{eq19}),
we obtain the metric functions as
\begin{equation}
\label{eq20} A(t) = (n \ell t + c_{1})^{\frac{3m}{n(m + 1)}} \;,
\end{equation}
\begin{equation}
\label{eq21} B(t) = c_{4}(n \ell t + c_{1})^{\frac{3m}{n(m + 1)}}
\exp{\left[\frac{(m+1)^{2}}{2\ell ^{2}M}(n\ell t +
c_{1})^{\frac{6(m-1)}{n(m+1)}+2 }+\frac{c_{3} }{ \ell (n-3)} (n\ell
t + c_{1})^{\frac{n - 3}{n} }\right]} \;,
\end{equation}
\begin{equation}
\label{eq22} C(t) = c_{4}^{-1}(n \ell t + c_{1})^{\frac{3(1-m)}{n(m
+ 1)}} \exp{\left[-\frac{(m+1)^{2}}{2\ell ^{2}M}(n\ell t +
c_{1})^{\frac{6(m-1)}{n(m+1)}+2 }-\frac{c_{3} }{ \ell (n-3)} (n\ell
t + c_{1})^{\frac{n - 3}{n} }\right]} \;,
\end{equation}
where $M=(9m-3+mn+n)(3m-3+mn+n)$ and $\;n\neq3$.

In the special case $n=3$, we have
\begin{equation}
\label{eq20} A(t) = (3 \ell t + c_{1})^{\frac{m}{m + 1}} \;,
\end{equation}
\begin{equation}
\label{eq21} B(t) = c_{4}(3 \ell t + c_{1})^{\frac{3m\ell+c_{3}(m +
1)}{3\ell(m + 1)}} \exp{\left[\frac{(m+1)^{2}}{144\ell
^{2}m^{2}}(3\ell t + c_{1})^{\frac{4m}{m+1}}\right]} \;,
\end{equation}
\begin{equation}
\label{eq22} C(t) = c_{4}^{-1}(3 \ell t +
c_{1})^{\frac{3(1-m)\ell-c_{3}(m + 1)}{3\ell(m + 1)}}
\exp{\left[-\frac{(m+1)^{2}}{144\ell ^{2}m^{2}}(3\ell t +
c_{1})^{\frac{4m}{m+1} }\right]} \;,
\end{equation}

Thus, the metric (\ref{eq1}) is completely determined.

The expressions for the isotropic pressure ($p$), the proper energy
density ($\rho$), the string tension ($\lambda$) and the particle
density ($\rho_{p}$) for the above model are obtained as
\begin{eqnarray}
  p &=& \frac{3\ell^{2}[n(m+1)- 3(m^{2}-m+1)]}{(m + 1)^{2}}
(n \ell t + c_{1})^{-2}+\frac{3(m+1)(n+1)}{4(9m-3+mn+n)}(n\ell t +
c_{1})^{-\frac{6(1 - m)} {n(m + 1)}} \nonumber\\
& & +\frac{3c_{3}\ell(1-2m)}{m+1}(n\ell t +c_{1})^{-\frac{3}{n}-1}-
\frac{(m+1)^{2}}{\ell^{2}(9m-3+mn+n)^{2}}(n\ell t +
c_{1})^{-\left[\frac{12(1 - m)} {n(m + 1)}-2\right]}\nonumber\\
& &- \frac{2c_{3}(m+1)}{\ell(9m-3+mn+n)}(n\ell t +
c_{1})^{-\left[\frac{3(3 - m)} {n(m + 1)}-1\right]}- c_{3}^{2}(n\ell
t + c_{1})^{-\frac{6}{n}},
\end{eqnarray}
\begin{eqnarray}
\rho &=& \frac{9\ell^{2}m(2-m)}{(m + 1)^{2}} (n \ell t +
c_{1})^{-2}+\frac{15-33m-mn-n}{4(9m-3+mn+n)}(n\ell t +
c_{1})^{-\frac{6(1 - m)} {n(m + 1)}}  \nonumber\\
& &+\frac{3c_{3}\ell(1-2m)}{m+1}(n\ell t +c_{1})^{-\frac{3}{n}-1}-
\frac{(m+1)^{2}}{\ell^{2}(9m-3+mn+n)^{2}}(n\ell t +
c_{1})^{-\left[\frac{12(1 - m)} {n(m + 1)}-2\right]} \nonumber\\
& &- \frac{2c_{3}(m+1)}{\ell(9m-3+mn+n)}(n\ell t+
c_{1})^{-\left[\frac{3(3 - m)} {n(m + 1)}-1\right]}- c_{3}^{2}(n\ell
t + c_{1})^{-\frac{6}{n}},
\end{eqnarray}

\begin{equation}
\label{eq26} \lambda = \frac{3\ell^{2}(2m-1)(3 - n)}{m + 1}(n\ell t
+ c_{1})^{-2} + 2(n\ell t + c_{1})^{-\frac{6(1 - m)}{n(m + 1)}} \;,
\end{equation}
\begin{eqnarray}
\rho_{p} &=& \frac{3\ell^{2}[3m(2-m))+(3-n)(1-2m)(m+1)]}{(m +
1)^{2}} (n \ell t +
c_{1})^{-2}+\frac{3(13-35m-3mn-3n)}{4(9m-3+mn+n)}(n\ell t +
c_{1})^{-\frac{6(1 - m)} {n(m + 1)}}  \nonumber\\
& &+\frac{3c_{3}\ell(1-2m)}{m+1}(n\ell t +c_{1})^{-\frac{3}{n}-1}-
\frac{(m+1)^{2}}{\ell^{2}(9m-3+mn+n)^{2}}(n\ell t +
c_{1})^{-\left[\frac{12(1 - m)} {n(m + 1)}-2\right]} \nonumber\\
& &- \frac{2c_{3}(m+1)}{\ell(9m-3+mn+n)}(n\ell t+
c_{1})^{-\left[\frac{3(3 - m)} {n(m + 1)}-1\right]}- c_{3}^{2}(n\ell
t + c_{1})^{-\frac{6}{n}}.
\end{eqnarray}

The above solutions satisfy the energy conservation equation
(\ref{eq11}) identically, as expected.

\begin{figure}[htbp]
\centering
\includegraphics[height=6cm,angle=0]{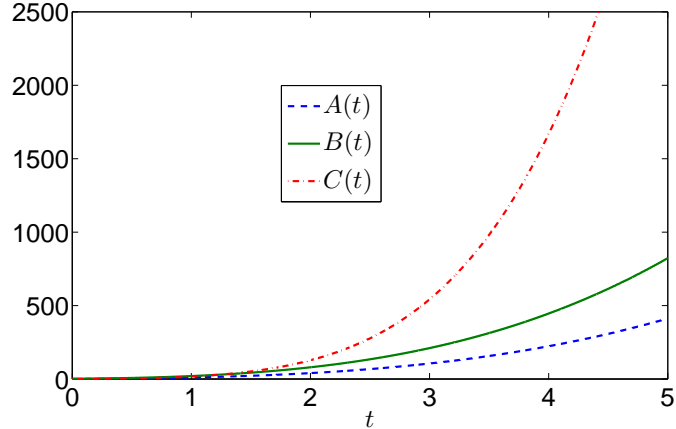}
\caption{Scale factors vs time with $n = 0.5$, $m = 0.4$, $\ell =
2$, $c_{1} = c_{3} = 1$ and $c_{4} = 2$.}
\end{figure}

We observe that all the parameters diverge at $t=-c_{1}/n\ell$.
Therefore, the model has a singularity at $t=-c_{1}/n\ell$, which
can be shifted to $t=0$ by choosing $c_{1}=0$. This singularity is
of Point Type as all the scale factors vanish at $t=-c_{1}/n\ell$.
The cosmological evolution of Bianchi-II space-time is expansionary
since all the scale factors monotonically increase with time (see,
Fig.1). So, the Universe starts expanding with a big bang
singularity in the derived model. The parameters $p$, $\rho$,
$\rho_{p}$ and $\lambda$ start off with extremely large values. In
particular, the large values of $\rho_{p}$ and $\lambda$ in the
beginning suggest that strings dominate the early Universe. For
sufficiently large times, $\rho_{p}$ and $\lambda$ become
negligible. Therefore, the strings disappear from the Universe for
larger times. That is why, the strings are not observable in the
present Universe.

The rates of expansion in the direction of $x$, $y$ and $z$ are
given by
\begin{equation}
\label{eq28} H_{x} = \frac{\dot{A}}{A} = \frac{3m\ell}{m+1}(n\ell t+c_{1})^{-1}\; ,
\end{equation}
\begin{equation}
\label{eq29} H_{y} = \frac{\dot{B}}{B} = \frac{3m\ell}{m+1}(n\ell t
+ c_{1})^{-1} +\frac{m+1}{\ell (9m-3+mn+n)}(n\ell t +
c_{1})^{-\frac{6(1 - m)} {n(m + 1)}+1}+ c_{3} (n\ell t +
c_{1})^{-\frac{3}{n}} \;,
\end{equation}
\begin{equation}
\label{eq30} H_{z} = \frac{\dot{C}}{C}=\frac{3(1-m)\ell}{m+1}(n\ell
t + c_{1})^{-1} -\frac{m+1}{\ell (9m-3+mn+n)}(n\ell t +
c_{1})^{-\frac{6(1 - m)} {n(m + 1)}+1}- c_{3} (n\ell t +
c_{1})^{-\frac{3}{n}} .
\end{equation}
The average Hubble's parameter, expansion scalar and shear of the
model are, respectively given by
\begin{equation}
\label{eq31} H = \frac{1}{3}(H_{x}+H_{y}+H_{z})=\ell(n\ell t +
c_{1})^{-1} ,
\end{equation}
\begin{equation}
\label{eq32} \theta = 3H=3\ell(n\ell t + c_{1})^{-1} ,
\end{equation}
\begin{eqnarray}
\sigma^{2} &=&
\frac{1}{6}\left[(H_{x}-H_{y})^{2}+(H_{y}-H_{z})^{2}+(H_{z}-H_{x})^{2}\right]\nonumber\\
&=&\frac{3\ell^{2}(2m-1)^{2}}{(m+1)^{2}} (n \ell t +
c_{1})^{-2}-\frac{15-33m-mn-n}{4(9m-3+mn+n)}(n\ell t +
c_{1})^{-\frac{6(1 - m)} {n(m + 1)}}  \nonumber\\
& &-\frac{3c_{3}\ell(1-2m)}{m+1}(n\ell t +c_{1})^{-\frac{3}{n}-1}+
\frac{(m+1)^{2}}{\ell^{2}(9m-3+mn+n)^{2}}(n\ell t +
c_{1})^{-\left[\frac{12(1 - m)} {n(m + 1)}-2\right]} \nonumber\\
& &+ \frac{2c_{3}(m+1)}{\ell(9m-3+mn+n)}(n\ell t+
c_{1})^{-\left[\frac{3(3 - m)} {n(m + 1)}-1\right]}+ c_{3}^{2}(n\ell
t + c_{1})^{-\frac{6}{n}}.
\end{eqnarray}

The spatial volume ($V$) and anisotropy parameter $(\bar{A})$ are found to be
\begin{equation}
\label{eq34} V = ABC=(n\ell t + c_{1})^{\frac{3}{n}},
\end{equation}

\begin{eqnarray}
\bar{A} &=& \frac{2\sigma^{2}}{3H^{2}}=\frac{2(2m-1)^{2}}{(m+1)^{2}}
-\frac{15-33m-mn-n}{6\ell ^{2}(9m-3+mn+n)}(n\ell t +
c_{1})^{-\frac{6(1 - m)} {n(m + 1)}+2}  \nonumber\\
& &-\frac{2c_{3}(1-2m)}{\ell(m+1)}(n\ell t +c_{1})^{-\frac{3}{n}+1}+
\frac{2(m+1)^{2}}{3\ell^{4}(9m-3+mn+n)^{2}}(n\ell t +
c_{1})^{-\left[\frac{12(1 - m)} {n(m + 1)}-4\right]} \nonumber\\
& &+ \frac{4c_{3}(m+1)}{3\ell^{3}(9m-3+mn+n)}(n\ell t+
c_{1})^{-\left[\frac{3(3 - m)} {n(m + 1)}-3\right]}+
\frac{2c_{3}^{2}}{3\ell^{2}}(n\ell t + c_{1})^{-\frac{6}{n}+2}.
\end{eqnarray}

The value of DP ($q$) is found to be
\begin{equation}\label{28}
q=-1+\frac{\dot{H}}{H^{2}}=n-1,
\end{equation}
which is a constant. A positive sign of $q$, i.e., $n>1$ corresponds
to the standard decelerating model whereas the negative sign of $q$,
i.e., $0< n<1$ indicates acceleration. The expansion of the Universe
at a constant rate corresponds to $n=1$, i.e., $q=0$. Also, recent
observations of SN Ia \cite{1a}-\cite{25a} reveal that the present
Universe is accelerating and value of DP lies somewhere in the range
$-1<q< 0.$ It follows that in the derived model, one can choose the
values of DP consistent with the observations.

From the above results, it can be seen that the spatial volume is
zero at $t=-c_{1}/n\ell$, and it increases with the cosmic time $t$.
The parameters $H_{x}$, $H_{y}$, $H_{z}$, $H$, $\theta$ and $\sigma$
diverge at the initial singularity. These parameters decrease with
the evolution of Universe, and finally drop to zero at late times
provided $m<1$. The mean anisotropy parameter asymptotically
approaches to $\frac{2(2m-1)^{2}}{(m+1)^{2}}$ for $n < 3$ and $m<1$.
Thus, the dynamics of the mean anisotropy parameter depends on the
values of $n$ and $m$. The model does not approach isotropy provided
$m\neq0.5$ as may be observed from Fig. 1. In case $m=0.5$, at late
times, the directional scale factors vary as
\[A(t)\approx(n \ell t + c_{1})^{\frac{1}{n}},
\;\;B(t)\approx(n \ell t + c_{1})^{\frac{1}{n}},\;\;C(t)\approx(n
\ell t + c_{1})^{\frac{1}{n}}.\]

Therefore, isotropy is achieved in the derived model for $m=0.5$.
Since the present-day Universe is isotropic, we consider $m=0.5$ in
the remaining discussion of the model.

For $n < 1$, the model is accelerating whereas for $n > 1$ it goes
to decelerating phase. In what follows, we compare the two modes of
evolution through graphical analysis of various parameters. We have
chosen $n=2$, i.e., $q=1$ to describe the decelerating phase while
the accelerating mode has been accounted by choosing $n=0.5$, i.e.,
$q=-0.5$. The other constants are chosen as $\ell=6$, $c_{1} = 1$,
$c_{3} = 1$, $m = 0.5$.

\begin{figure}[htbp]
\centering
\includegraphics[height=6cm,angle=0]{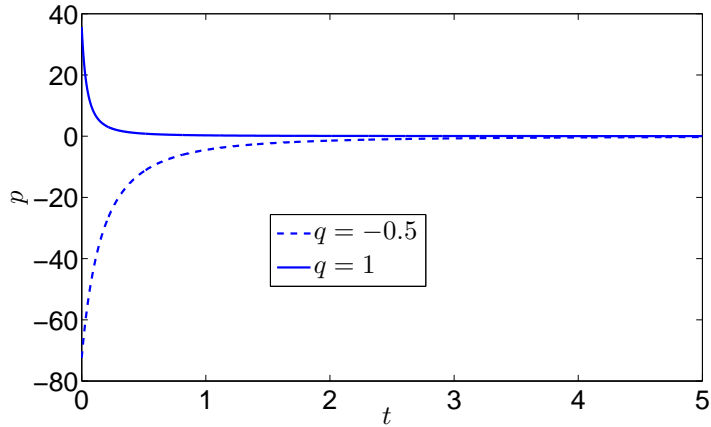}
\caption{Pressure vs time.}
\end{figure}

\begin{figure}[htbp]
\centering
\includegraphics[height=6cm,angle=0]{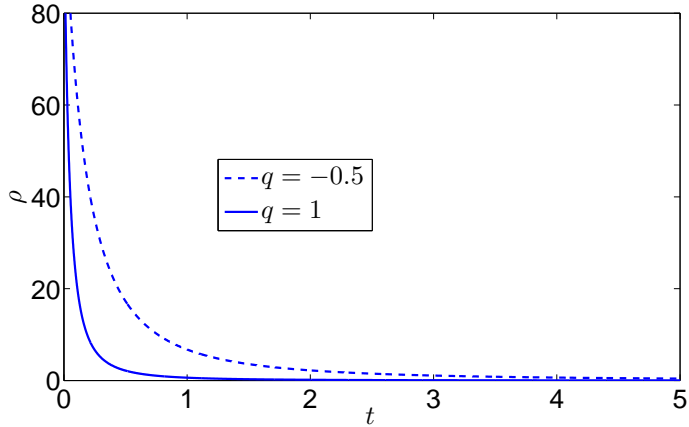}
\caption{Rest energy density vs time.}
\end{figure}
\begin{figure}[htbp]
\centering
\includegraphics[height=6cm,angle=0]{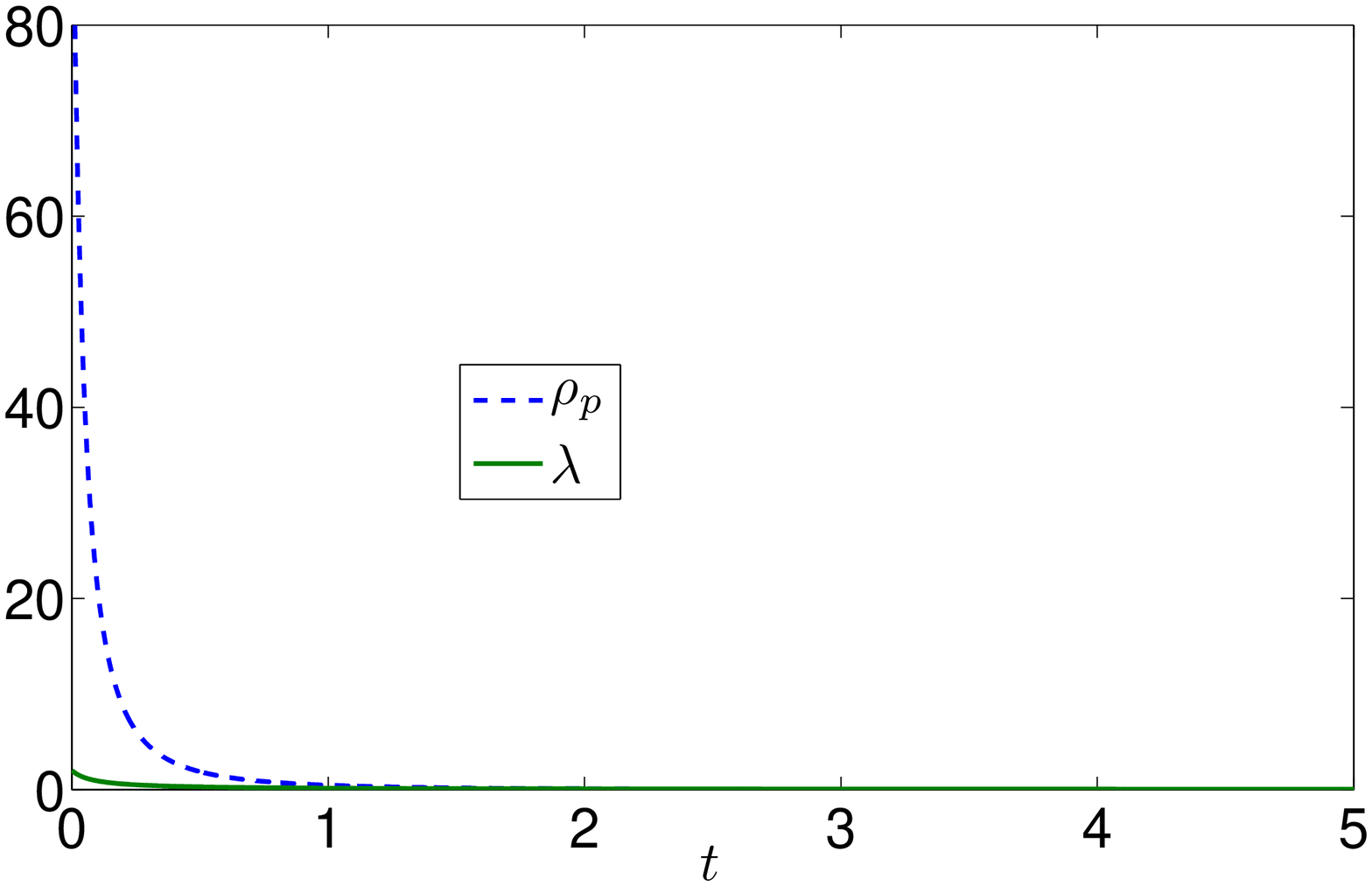}
\caption{Particle energy density and string tension vs time for
$q=1$.}
\end{figure}

\begin{figure}[htbp]
\centering
\includegraphics[height=6cm,angle=0]{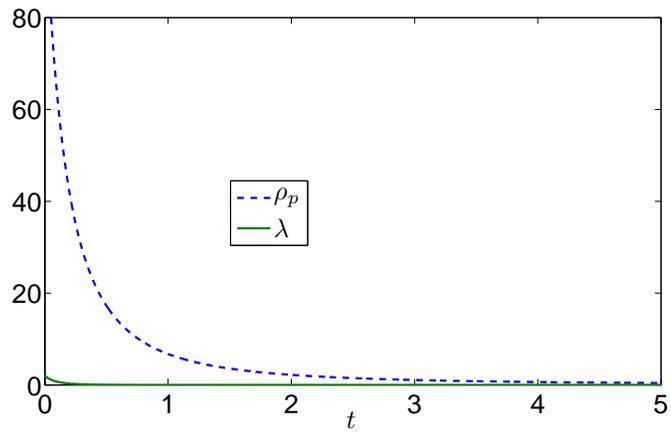}
\caption{Particle energy density and string tension vs time for
$q=-0.5$.}
\end{figure}

Fig. 2 depicts the variation of pressure versus time in the two
modes of evolution of the Universe. We observe that the pressure is
positive in the decelerating Universe which decreases with the
evolution of the Universe. But in the accelerating phase, negative
pressure dominates the Universe, as expected. In both cases, the
pressure becomes negligible at late times.

The rest energy density has been graphed versus time in Fig. 3. It
is evident that the rest energy density remains positive in both
modes of evolution. However, it decreases more sharply with the
cosmic time in the decelerating Universe.

Fig. 4 and Fig. 5 show the behavior of particle energy density and
string tension versus time in the decelerating and accelerating
modes, respectively. We see that $\rho_{p}>\lambda$, i.e., the
particle energy density remains larger than the string tension
density during the cosmic expansion , especially in early Universe.
This shows that massive strings dominate the early Universe (see,
Refs. \cite{ref1,ref50}). Further, it is observed that for
sufficiently large times, $\rho_{p}$ and $\lambda$ tend to zero.
Therefore, the strings disappear from the Universe at late times.

According to Ref. \cite{ref50}, since there is no direct evidence of
strings in the present-day Universe, we are in general, interested
in constructing models of a Universe that evolves purely from the
era dominated by either geometric string or massive strings and ends
up in a particle dominated era with or without remnants of strings.
Therefore, the above model describes the evolution of the Universe
consistent with the present-day observations.

\begin{figure}[htbp]
\centering
\includegraphics[height=6cm,angle=0]{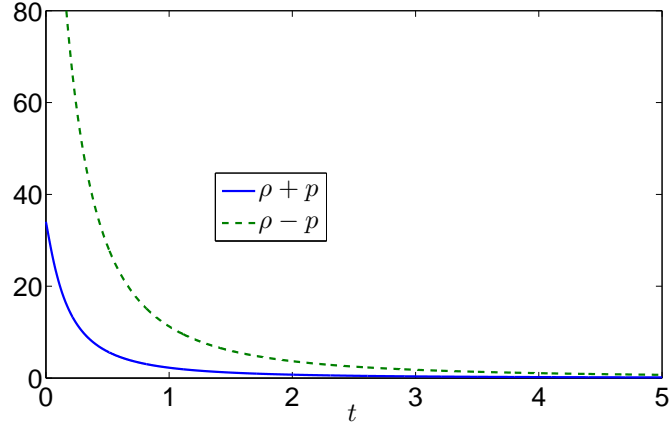}
\caption{$\rho+p$ and $\rho-p$ vs time for $q=-0.5$.}
\end{figure}

From Fig. 3 to Fig. 6, we observe the following:\\

(i) $\rho\geq0$ \\

(ii) $\rho_{p}\geq0$ \\

(iii) $\rho+p\geq0$ \\

(iv) $\rho-p\geq0$.\\

This shows that the weak and dominant energy conditions are
satisfied in the derived model.
\subsection{String Cosmology with Exponential-law}
Solving the equations (\ref{eq17}), (\ref{eq12}) and (\ref{eq19}),
we obtain the metric functions as
\begin{equation}
\label{eq39} A(t) =
c_{2}^{\frac{3m}{m+1}}\exp{\left(\frac{3m\ell}{m+1}t\right)}\; ,
\end{equation}
\begin{equation}
\label{eq40} B(t) =
c_{4}c_{2}^{\frac{3m}{m+1}}\exp{\left[\frac{3m\ell}{m+1}t
+\frac{c_{2}^{\frac{6(m-1)}{m+1}}(m+1)^{2}}{18\ell^{2}(3m-1)(m-1)}e^{\frac{6\ell(m-1)}{m+1}t}-
\frac{c_{3}} {3\ell c_{2}^{3}}e^{-3\ell t}\right]},
\end{equation}
\begin{equation}
\label{eq41} C(t) =
c_{4}^{-1}c_{2}^{\frac{3(1-m)}{m+1}}\exp{\left[\frac{3(1-m)\ell}{m+1}t
-\frac{c_{2}^{\frac{6(m-1)}{m+1}}(m+1)^{2}}{18\ell^{2}(3m-1)(m-1)}e^{\frac{6\ell(m-1)}{m+1}t}+
\frac{c_{3}} {3\ell c_{2}^{3}}e^{-3\ell t}\right]}.
\end{equation}

The expressions for the isotropic pressure, the proper energy
density, the string tension and the particle density for the derived
model are obtained as
\begin{eqnarray}
\label{eq43} p &=& -\frac{9\ell^{2}(m^{2}-m+1)}{(m + 1)^{2}}
+\frac{c_{2}^{\frac{6(m-1)}{m+1}}(m+1)}{4(3m-1)}e^{\frac{6\ell(m-1)}{m+1}t}
-\frac{2c_{3}c_{2}^{\frac{3(m-3)}{m+1}}(m+1)}{3\ell(3m-1)}e^{\frac{3\ell(m-3)}{m+1}t}
\nonumber\\
& &+\frac{3c_{3}(1-2m)} {c_{2}^{3}(m+1)}e^{-3\ell t}
-\frac{c_{2}^{\frac{12(m-1)}{m+1}}(m+1)^{2}}{9\ell^{2}(3m-1)^{2}}e^{\frac{12\ell(m-1)}{m+1}t}
-\frac{c_{3}^{2}}{c_{2}^{6}}e^{-6\ell t},
\end{eqnarray}
\begin{eqnarray}\label{eq44}
\rho &=& \frac{9\ell^{2}m(2-m)}{(m + 1)^{2}}
+\frac{c_{2}^{\frac{6(m-1)}{m+1}}(5-11m)}{4(3m-1)}e^{\frac{6\ell(m-1)}{m+1}t}
-\frac{2c_{3}c_{2}^{\frac{3(m-3)}{m+1}}(m+1)}{3\ell(3m-1)}e^{\frac{3\ell(m-3)}{m+1}t}
\nonumber\\
& &+\frac{3c_{3}(1-2m)} {c_{2}^{3}(m+1)}e^{-3\ell t}
-\frac{c_{2}^{\frac{12(m-1)}{m+1}}(m+1)^{2}}{9\ell^{2}(3m-1)^{2}}e^{\frac{12\ell(m-1)}{m+1}t}
-\frac{c_{3}^{2}}{c_{2}^{6}}e^{-6\ell t}
\end{eqnarray}
\begin{equation}
\label{eq45} \lambda = \frac{9\ell^{2}(2m-1)}{m +
1}+2c_{2}^{\frac{6(m - 1)}{m + 1}} e^{\frac{6\ell(m-1)}{m+1}t},
\end{equation}
\begin{eqnarray}\label{eq44}
\rho_{p} &=& \frac{9\ell^{2}(1+m-3m^{2})}{(m + 1)^{2}}
+\frac{c_{2}^{\frac{6(m-1)}{m+1}}(13-25m)}{4(3m-1)}e^{\frac{6\ell(m-1)}{m+1}t}
-\frac{2c_{3}c_{2}^{\frac{3(m-3)}{m+1}}(m+1)}{3\ell(3m-1)}e^{\frac{3\ell(m-3)}{m+1}t}\nonumber\\
& &+\frac{3c_{3}(1-2m)} {c_{2}^{3}(m+1)}e^{-3\ell t}
-\frac{c_{2}^{\frac{12(m-1)}{m+1}}(m+1)^{2}}{9\ell^{2}(3m-1)^{2}}e^{\frac{12\ell(m-1)}{m+1}t}
-\frac{c_{3}^{2}}{c_{2}^{6}}e^{-6\ell t}.
\end{eqnarray}

The energy conservation equation (\ref{eq11}) is satisfied identically by the above solutions, as expected.

The directional Hubble's parameters are given by
\begin{equation}
\label{eq51} H_{x} =  \frac{3m\ell}{m + 1},
\end{equation}
\begin{equation}
\label{eq52} H_{y} =  \frac{3m\ell}{m+1}
+\frac{c_{2}^{\frac{6(m-1)}{m+1}}(m+1)}{3\ell(3m-1)}e^{\frac{6\ell(m-1)}{m+1}t}+
\frac{c_{3}} {c_{2}^{3}}e^{-3\ell t},
\end{equation}
\begin{equation}
\label{eq53} H_{z} =  \frac{3\ell(1-m)}{m+1}
-\frac{c_{2}^{\frac{6(m-1)}{m+1}}(m+1)}{3\ell(3m-1)}e^{\frac{6\ell(m-1)}{m+1}t}-
\frac{c_{3}} {c_{2}^{3}}e^{-3\ell t}.
\end{equation}

Hence the average generalized Hubble's parameter is given by
\begin{equation}
\label{eq54} H = \ell.
\end{equation}

From equations (\ref{eq51})-(\ref{eq54}), we observe that the
directional Hubble's parameters are time dependent while the average
Hubble's parameter is constant.

The expressions for kinematical parameters, i.e., the scalar of
expansion, shear scalar, the spatial volume, average anisotropy
parameter and deceleration parameter for the derived model are given
by
\begin{equation}
\label{eq47} \theta = 3 \ell,
\end{equation}

\begin{eqnarray}\label{eq48}
\sigma^{2} &=& \frac{3\ell^{2}(2m-1)^{2}}{(m + 1)^{2}}
-\frac{c_{2}^{\frac{6(m-1)}{m+1}}(5-11m)}{4(3m-1)}e^{\frac{6\ell(m-1)}{m+1}t}
+\frac{2c_{3}c_{2}^{\frac{3(m-3)}{m+1}}(m+1)}{3\ell(3m-1)}e^{\frac{3\ell(m-3)}{m+1}t}
\nonumber\\
& &-\frac{3c_{3}(1-2m)} {c_{2}^{3}(m+1)}e^{-3\ell t}
+\frac{c_{2}^{\frac{12(m-1)}{m+1}}(m+1)^{2}}{9\ell^{2}(3m-1)^{2}}e^{\frac{12\ell(m-1)}{m+1}t}
+\frac{c_{3}^{2}}{c_{2}^{6}}e^{-6\ell t},
\end{eqnarray}

\begin{equation}
\label{eq49} V = c_{2}^{3}e^{3\ell t},
\end{equation}
\begin{eqnarray}\label{eq48}
\bar{A} &=& \frac{2(2m-1)^{2}}{(m + 1)^{2}}
-\frac{c_{2}^{\frac{6(m-1)}{m+1}}(5-11m)}{6\ell^{2}(3m-1)}e^{\frac{6\ell(m-1)}{m+1}t}
+\frac{4c_{3}c_{2}^{\frac{3(m-3)}{m+1}}(m+1)}{9\ell^{3}(3m-1)}e^{\frac{3\ell(m-3)}{m+1}t}
\nonumber\\
& &-\frac{2c_{3}(1-2m)} {\ell^{2}c_{2}^{3}(m+1)}e^{-3\ell t}
+\frac{2c_{2}^{\frac{12(m-1)}{m+1}}(m+1)^{2}}{27\ell^{4}(3m-1)^{2}}e^{\frac{12\ell(m-1)}{m+1}t}
+\frac{2c_{3}^{2}}{3\ell^{2}c_{2}^{6}}e^{-6\ell t},
\end{eqnarray}
\begin{equation}
\label{eq50} q = - 1.
\end{equation}

Recent observations of SN Ia \cite{1a}-\cite{25a} suggest that the
Universe is accelerating in its present state of evolution. It is
believed that the way Universe is accelerating presently; it will
expand at the fastest possible rate in future and forever. For
$n=0$, we get \textbf{ $q=-1$ }; incidentally this value of DP leads
to $dH/dt=0$, which implies the greatest value of Hubble's parameter
and the fastest rate of expansion of the Universe. Therefore, the
derived model can be utilized to describe the dynamics of the late
time evolution of the actual Universe. So, in what follows, we
emphasize upon the late time behavior of the derived model. At late
times, we find

\begin{equation}
p \approx -\frac{9\ell^{2}(m^{2}-m+1)}{(m + 1)^{2}},
\end{equation}
\begin{equation}
\rho \approx \frac{9\ell^{2}m(2-m)}{(m + 1)^{2}} ,
\end{equation}
\begin{equation}
\lambda \approx \frac{9\ell^{2}(2m-1)}{m + 1},
\end{equation}
\begin{equation}
\rho_{p} \approx \frac{9\ell^{2}(1+m-3m^{2})}{(m + 1)^{2}} ,
\end{equation}
\begin{equation}
\sigma^{2}\approx \frac{3\ell^{2}(2m-1)^{2}}{(m + 1)^{2}},
\end{equation}
\begin{equation}
\bar{A} \approx \frac{2(2m-1)^{2}}{(m + 1)^{2}}.
\end{equation}

In particular, for $m=\frac{1}{2}$, we have

$$p=-\rho,$$
$$\lambda\approx0,$$
$$\rho_{p} \approx 3\ell^{2},$$
$$\sigma^{2}\approx0,$$
$$\bar{A}\approx0.$$

This shows that vacuum energy dominates the Universe at late times,
which is consistent with the observations. Strings disappear and the
Universe evolves with constant particle energy density. The shear
and anisotropy parameter become negligible. So the Universe becomes
isotropic.
\section{Concluding Remarks}
In this paper, a spatially homogeneous and anisotropic Bianchi-II
space-time representing massive strings in general relativity has
been studied. The main features of the work are as follows:
\begin{itemize}
\item The models are based on exact solutions of the Einstein's field equations for the anisotropic Bianchi-II
space-time filled with massive strings.
\item The singular model ($n\neq 0$) seems to describe the dynamics of
    Universe from big bang to the present epoch while the
    non-singular model ($n=0$) seems reasonable to project dynamics of
    future Universe.
\item In the derived models, $m=\frac{1}{2}$ turns out to be a condition of isotropy.
    \item In the present models, the
weak and dominant energy conditions are satisfied, which in turn
imply that the derived models are physically realistic.
\item The singular model presents the dynamics of strings in the accelerating
and decelerating modes of evolution of the Universe. It has been
found that massive strings dominate the Universe, which eventually
disappear from the Universe for sufficiently large times. This is in
agreement with the astronomical observations.
\item The non-singular model for $m=\frac{1}{2}$ predicts a Universe dominated by
vacuum energy, which is consistent with the predictions of current
observations.
\end{itemize}
\noindent\textbf{Acknowledgements:}\\
The author is thankful to the anonymous referee for valuable
comments on this manuscript.

\end{document}